\documentclass[aps,pre,twocolumn,groupedaddress]{revtex4-1}

\usepackage{graphics}
\usepackage{graphicx}
\usepackage{float}
\usepackage{amssymb}
\usepackage{bbm}

\begin{document}

\title{Mesoscopic approach to granular crystal dynamics}

\author{Marcial Gonzalez}
\author{Jinkyu Yang}
\author{Chiara Daraio}
\author{Michael Ortiz}
\affiliation{Graduate Aerospace Laboratories, California Institute of Technology,
             Pasadena, CA 91125, USA}

\date{\today}

\begin{abstract}
We present a mesoscopic approach to granular crystal dynamics, which comprises a three-dimensional finite-element model and a one-dimensional regularized contact model. The approach investigates the role of vibrational-energy trapping effects in the dynamic behavior of one-dimensional chains of particles in contact (i.e., granular crystals), under small to moderate impact velocities. The only inputs of the models are the geometry and the elastic material properties of the individual particles that form the system. We present detailed verification results and validate the model comparing its predictions with experimental data. This approach provides a physically sound, first-principle description of dissipative losses in granular systems.
\end{abstract}

\maketitle

\section{Introduction}
Granular crystals, or highly-packed granular lattices, are strongly nonlinear systems that exhibit unique dynamics. In particular, a one-dimensional chain of elastic spherical beads supports the formation and propagation of compact solitary waves as the result of Hertzian nonlinear contact interactions between the particles in the system \cite{Nesterenko-2001}. The width of these waves is independent of their amplitude $F_\mathrm{m}$---and their wave speed follows $V_s \propto F_\mathrm{m}^{1/6}$. This is in contrast to disordered granular media where the nature of the system is additionally driven by frictional and rotational dynamics. The dynamic response of granular crystals can be tuned by varying the particles' material and size, or by the application of precompression to the system \cite{Nesterenko-2001,Coste-1997,Daraio-2006a}. Over the last decade, the response of these systems has drawn considerable attention and many potential applications have been studied, such as vibration mitigation \cite{Boechler-2010}, shock and energy absorption \cite{Herbold-2007,Daraio-2006b,Hong-2005}, actuators \cite{Khatri-2008}, and sound focusing devices \cite{Spadoni-2010}.

A typical experimental setup commonly used in the study of granular crystal dynamics consists of a monodispersed one-dimensional chain of spherical beads guided by straight rails and impacted by a striker. Selected beads are instrumented with calibrated piezosensors \cite{Daraio-2005a} that allow measurements of the forces acting inside the particles. These measurements confirmed the formation of compact solitary waves, and reported the presence of a characteristic amplitude decay, as the solitary waves propagate through the system (cf., e.g., \cite{Carretero-2009}).

A particle mechanics approach has been extensively used in the literature to model granular crystal dynamics. In this approach, dissipative effects are neglected and the interaction between beads is assumed to follow Hertz law \cite{Nesterenko-2001,Daraio-2006a,Daraio-2006b,Hong-2005,Khatri-2008,Spadoni-2010,Daraio-2005a,Porter-2009}. This model does indeed predict the formation of solitary waves of constant width that travel along the chain with a speed $V_s \propto F_\mathrm{m}^{1/6}$, but it does not capture the experimentally-observed decay of the force. This discrepancy has recently motivated the inclusion of dissipative effects in the model, such as friction, plasticity, viscoelasticity, and viscous drag \cite{Carretero-2009,Coste-1997,Herbold-2007,Rosas-2007,Sen-2008,Vergara-2010}. A fundamental challenge in modeling dissipative losses in granular crystals is the ability to capture quantitatively the amplitude decay and the wave shape variations observed experimentally, as the wave travels through the systems. This was done in \cite{Carretero-2009} modeling dissipation in the form of an empirical discrete Laplacian in the velocities, and relying on two phenomenological parameters obtained from best fitting experimental observations. In order to design engineering devices and materials that exploit the unique wave dynamics of granular crystals, it is important to address the role of dissipation using a first-principle description that relies only on the knowledge of the particles' geometry and materials properties.

This work is concerned with the formulation of first-principles predictive models of granular crystal dynamics. In particular, we investigate the role of vibrational-energy trapping effects in the overall dynamic behavior of one-dimensional chains under small to moderate impact velocities. To this end, we follow a mesoscopic approach comprised by two models whose only inputs are the geometry and elastic material properties of each particle in the system. The first model resolves the fine mesoscale structure of dynamic collisions and explicitly accounts for the vibrational energy retained in each bead as the solitary wave propagates along the chain. For small to moderate impact velocities, the model is conservative and the inclusion of permanent energy losses (such as plasticity and viscoelasticity) is not required. We achieve these properties by abandoning the particle mechanics approach and adopting a three-dimensional finite-element model (i.e., a dynamic contact problem of three-dimensional deformable elastic bodies that interact with one another over time is solved).

The second model proposed in this work is a one-dimensional regularized contact model where the vibrational energy that remains trapped after impact is subsumed under the concept of a coefficient of restitution. For small to moderate impact velocities, the variation of this coefficient with the impact velocity is a geometry and material dependent property that solely accounts for mesoscopic dynamic effects and that can be obtained from an experimental (cf., e.g., \cite{Zhang-2009}) or numerical (cf., \cite{Giese-1996,Aspelmeier-1998} for elastic rods, and \cite{Gerl-1999} for elastic disks) campaign of head-on collisions. This model is inherently energy-consistent and momentum-preserving.

The concept of translational kinetic energy being transferred to internal vibrational degrees of freedom during the collision of elastic particles dates back to Rayleigh \cite{Rayleigh-1906}. For the case of two slowly colliding spheres, Rayleigh restricted attention to the fundamental vibrational mode and estimated that the coefficient of restitution is a function of the impact velocity and does not vanish for identical spheres. However, an elastic sphere has a rich family of toroidal and spheroidal vibrational modes \cite{Lamb-1881} and, therefore, it is expected that a fraction of energy is also stored in modes higher than the fundamental mode. Zippelius and co-workers have analytically shown that this is indeed the case for lower dimensional objects, such as rods and disks, and they developed microscopic models for the coefficient of restitution \cite{Giese-1996,Aspelmeier-1998,Gerl-1999}. Unfortunately, the extension of their approach to elastic spherical particles renders the problem intractable. Hence, a three-dimensional finite-element model is proposed in this work.

\section{Mesoscopic approach to granular crystal dynamics}

\textbf{\emph{Three-dimensional finite-element model}}.
The first model presented in this work makes use of a three-dimensional finite-element mesh of the chain of beads. The contact constraint between beads is enforced using a penalty energy function \cite{Gonzalez-2010a,Gonzalez-2010b}. The trajectory of the elastodynamic problem naturally follows from Hamilton's principle
\begin{equation}
\label{Eqn-MotionFEM}
\textstyle
    0
    =
    \delta \int \left[ \text{KE}(\dot{\mathbf{q}}) - V(\mathbf{q}) - I_{\mathcal{C}}(\mathbf{q}) \right]~\mathrm{d}t
\end{equation}
where the generalized coordinates $\mathbf{q}(t)$ are the coordinates of the finite-element nodes in the deformed configuration, $\text{KE}(\dot{\mathbf{q}})=\frac{1}{2} \dot{\mathbf{q}}^T \mathbf{M} \dot{\mathbf{q}}$ is the kinetic energy, $\mathbf{M}$ is the mass matrix, $V(\mathbf{q})$ is the elastic energy, and $I_{\mathcal{C}}(\mathbf{q})$ is the impenetrability or contact constraint. The impenetrability condition is enforced by penalizing the intersection between pairs of element faces (see \cite{Gonzalez-2010a} and references therein). The elastic behavior of the beads is described by the strain-energy density of a neo-Hookean solid extended to the compressible range \cite{Gonzalez-2010a}. The resulting dynamical system is then momentum and energy preserving, and its trajectories are obtained by numerical time integration of (\ref{Eqn-MotionFEM}). The multiple time scales---that coexist in time and space---and the complex dynamic contact of this three-dimensional system pose great challenges for numerical time integrators. We address these challenges with energy-stepping integrators which have proven to be effective and efficient in solving the dynamic behavior of a chain of elastic beads. These time integrators are energy-momentum conserving, symplectic, and convergent with automatic and asynchronous selection of the time step \cite{Gonzalez-2010a,Gonzalez-2010b}.

Of particular interest to granular crystal dynamics are conservation of total energy---$\text{KE}(\dot{\mathbf{q}})+V(\mathbf{q})$---and linear momentum. The kinetic energy and the linear momentum of the system can also be expressed as the contribution of all individual particles, i.e., $\text{KE}_k$ and $\mathbf{J}_k= M_k \langle\dot{\mathbf{q}}\rangle_k$, where $M_k$ and $\langle\dot{\mathbf{q}}\rangle_k$ are the mass and the mean velocity of particle $k$. Furthermore, the additive decomposition of the kinetic energy into a rigid-motion component---$\langle\text{KE}\rangle_k = \frac{1}{2} M_k \|\langle\dot{\mathbf{q}}\rangle_k\|^2$---and a vibrational contribution---$\text{VKE}_k=\text{KE}_k - \langle\text{KE}\rangle_k$---will allow for investigating the role of energy-trapping effects in granular crystals.

For the purpose of validating the model, it is essential to identify and compute forces that resemble those measured by embedded piezosensors in experimental setups, that is, following \cite{Daraio-2005a}, an averaged value $F_k^\mathrm{av}$ of the contact forces acting on the particle. Any particle $k$, except for the striker, has a contact force $\mathbf{F}_k^\mathrm{l}$ acting on its left and a contact force $\mathbf{F}_k^\mathrm{r}$ acting on its right; thus, the averaged force simplifies to $F_k^\mathrm{av} = \frac{1}{2} \left( \mathbf{F}_k^\mathrm{l} - \mathbf{F}_k^\mathrm{r} \right) \cdot \hat{\mathbf{n}}$, where $\hat{\mathbf{n}}$ is a unit vector aligned with the chain of beads. According to Newton's laws, the external force acting on particle $k$ is
\[
    \mathbf{F}_k^\mathrm{ext}
    =
    \mathbf{F}_k^\mathrm{l} + \mathbf{F}_k^\mathrm{r}
    =
    \frac{\mathrm{d}}{\mathrm{d}t}\mathbf{J}_k
    =
    M_k \frac{\mathrm{d}}{\mathrm{d}t}\langle\dot{\mathbf{q}}\rangle_k
\]
and, for a chain numbered from left to right, $\mathbf{F}_k^\mathrm{r} =-\mathbf{F}_{k+1}^\mathrm{l}$. Therefore, if the striker is referred as the first bead, $F_k^\mathrm{av}$ can be readily obtained from all $\mathbf{F}_{j\leq k}^\mathrm{ext}$.

\textbf{\emph{One-dimensional regularized contact model}}.
The second model presented in this work is motivated by a one-dimensional regularization of the three-dimensional contact problem previously presented. Then, for a one-dimensional chain of particles, we define the displacement $u_k(t) = [\langle \mathbf{q} \rangle_k(t) - \langle \mathbf{q} \rangle_k(0)] \cdot \hat{\mathbf{n}}$ and the velocity $\dot{u}_k(t) = \langle\dot{\mathbf{q}}\rangle_k(t) \cdot \hat{\mathbf{n}}$, where $k$ is the particle number--we assume other components of the displacement are zero. This intuitive dimensional reduction suggests recasting the three-dimensional Lagrangian system into a one-dimensional mechanical system with forcing. Thus, displacements $\mathbf{u}(t)$ are given by the Lagrange-d'Alembert principle, i.e.,
\begin{equation}
\label{Eqn-MotionForcing}
\textstyle
    0 = \delta \int \left[ \langle\text{KE}\rangle(\dot{\mathbf{u}}) - \bar{V}(\mathbf{u}) \right]~\mathrm{d}t
        +
        \int \mathbf{F}\left(\mathbf{u},\dot{\mathbf{u}}\right) \cdot \delta{\mathbf{u}}~\mathrm{d}t
\end{equation}
where $\mathbf{F}$ is a forcing term and $\bar{V}$ is a one-dimensional regularized potential. It is worth noting that the forcing term results in an effective dissipation of the vibrational energy retained in the particles during the collision.

We approximate the forcing term by a regularized contact model or compliant contact-force model. For direct central and frictionless impacts of two particles, Hunt and Crossley \cite{Hunt-1975} proposed a compliant normal-force model of the form
\[
    m (\ddot{u}_1 - \ddot{u}_2)
    =
    - \kappa \gamma^{n} - (\alpha \kappa \gamma^n) \dot{\gamma}
\]
where $\gamma = \max\{u_1 - u_2,0\}$ and $\dot{\gamma} = \dot{u}_1 - \dot{u}_2$ are the penetration depth and speed, $\alpha$ is a damping factor, $\kappa$ is a spring constant, and $m$ is the effective mass (i.e., $m^{-1} = M_1^{-1} + M_2^{-1}$). This regularized contact model falls squarely within the dimensional reduction proposed above, that is
\begin{eqnarray*}
    \bar{V}(\gamma) &= \frac{\kappa}{n+1}\gamma^{n+1}
    \\
    F(\gamma,\dot{\gamma}) &= -(\alpha \kappa \gamma^n) \dot{\gamma}
\end{eqnarray*}
Evidently, the potential energy is chosen in analogy with Hertz's theory, which is a good regularized model for the static contact problem of elastic bodies whose contact region remains small. Then, for heterogeneous pairs of linear elastic spheres, $n=3/2$ and the spring constant $\kappa$ is well-known (cf., e.g., \cite{Porter-2009}). The damping factor $\alpha$ is generally chosen to ensure that the energy dissipated during impact is consistent with the energy loss subsumed in a coefficient of restitution $e$.

Many researchers have proposed approximate \cite{Hunt-1975,Marhefka-1999,Lankarani-1990} and exact \cite{Gonthier-2004} relationships between $\alpha$ and $e$ (see, for example, \cite{Zhang-2009} for a detailed comparison of the predictions of these models with low speed impact measurements). In this work we adopt the exact solution proposed by Gonthier and co-workers \cite{Gonthier-2004}, i.e., the damping factor is determined from the implicit relation
\begin{equation}
\label{Eqn-Gonthier-1}
    \frac{1+\alpha v_\mathrm{i}}{1-\alpha v_\mathrm{i} e} = \exp\left(\alpha v_\mathrm{i}(1+e)\right)
\end{equation}
where $e$ and $v_\mathrm{i}$ are given \cite{Sln-Alpha}. It is worth noting that the subscript in $v_\mathrm{i}$ is not an index, it stands for the head-on impact velocity (i.e., the penetration speed at the start of the collision). This model is momentum-preserving and energy-consistent for all values of $e$. At low impact velocities, recent experimental data for steel \cite{Zhang-2009} suggest that the coefficient of restitution can be approximated by $e= 1 - c_1 v_\mathrm{i}^{c_2}$. Then, $\alpha$ is only a function of $v_\mathrm{i}$, and the empirical coefficients $c_1$, $c_2$ can be obtained from an experimental or numerical campaign of head-on collisions over a range of $v_\mathrm{i}$. It bears emphasis that, in this work, $e(v_\mathrm{i})$ accounts for the vibrational energy retained in the beads the during collision.

The application of Hunt-Crossley's regularized contact model to equations of motion (\ref{Eqn-MotionForcing}) results in
\begin{equation}
\label{Eqn-DissipativeParticleMechanics}
\textstyle
    M_k \ddot{u}_k
    =
    \kappa
    \left[
        \gamma_{k}^{3/2} (1+ \alpha_{k} \dot{\gamma}_{k})
        -
        \gamma_{k+1}^{3/2} (1+\alpha_{k+1} \dot{\gamma}_{k+1})
    \right]
\end{equation}
where $\gamma_k = \max\{u_k - u_{k+1},0\}$, $\dot{\gamma}_k = \dot{u}_k - \dot{u}_{k+1}$, and $\alpha_k = \alpha_k(v_{\mathrm{i},k})$ is given by Gonthier's energy-consistent model (\ref{Eqn-Gonthier-1}). The value for $v_{\mathrm{i},k}$ is approximated by the largest attained relative velocity between adjacent particles $k$ and $k+1$. Thus, the evolutionary equations for $v_{\mathrm{i},k}$ and its time derivative $a_{\mathrm{i},k} := v'_{\mathrm{i},k}$ are well-defined and given by:
\begin{enumerate}
  \item \emph{Loading/unloading conditions} ($\gamma_k > 0$): the irreversible nature of the percussion is captured by Kuhn-Tucker conditions (i.e., $a_{\mathrm{i},k} \geq 0$, $\dot{\gamma}_k - v_{\mathrm{i},k} \leq 0$, and $a_{\mathrm{i},k} (\dot{\gamma}_k - v_{\mathrm{i},k}) = 0$), together with a consistency condition if $\dot{\gamma}_k - v_{\mathrm{i},k} = 0$ (i.e., $a_{\mathrm{i},k} = \ddot{u}_k-\ddot{u}_{k+1}$ if $\ddot{u}_k-\ddot{u}_{k+1} \geq 0$; or $a_{\mathrm{i},k} = 0$ if $\ddot{u}_k-\ddot{u}_{k+1} < 0$).
  \item \emph{Touching/detaching conditions} ($\gamma_k = 0$): before and after the percussion of two beads, the value of $v_{\mathrm{i},k}$ may experience a jump given by $v_{\mathrm{i},k} = \dot{\gamma}_k$, at the first instant of contact ($\dot{\gamma}_k > 0$), or given by $v_{\mathrm{i},k} = 0$, after the collision ($\dot{\gamma}_k \leq 0$). Note that $a_{\mathrm{i},k}$ then exists almost everywhere.
\end{enumerate}
A distinguishing characteristic of the proposed model is that damping factors $\alpha_k$ are not assigned \emph{a priori} to each pair-interaction but directly follow from the integration of equations of motion (\ref{Eqn-DissipativeParticleMechanics}). This is achieved by additionally solving for the evolution in time of internal variables $v_{\mathrm{i},k}$ that account for the history-dependent nature of the dissipative process. In this respect, it is evident the similarity with models of plastic deformation---another path-dependent and irreversible phenomenon.

The treatment of gravitational effects falls squarely within the framework considered above and it is readily achieved by including a forcing term $\mathbf{F}^g$ in the equations of motion (\ref{Eqn-MotionForcing}), that is
\begin{equation}
    0 = \delta \int \left[ \langle\text{KE}\rangle(\dot{\mathbf{u}}) - \bar{V}(\mathbf{u}) \right]~\mathrm{d}t
        +
        \int \left[\mathbf{F}\left(\mathbf{u},\dot{\mathbf{u}}\right)+\mathbf{F}^g \right] \cdot \delta{\mathbf{u}}~\mathrm{d}t
    \label{Eqn-MotionForcingGravity}
\end{equation}
where $F^g_k = g M_k$ and $g$ is the gravitational constant. Finally, the model is integrated over time with variational integrators which accurately capture the energy behavior of forced mechanical systems \cite{Kane-2000}.

\section{Verification}

\begin{figure}[b]
\centering{
    \begin{tabular}{ccc}
    \includegraphics[width=0.455\columnwidth]{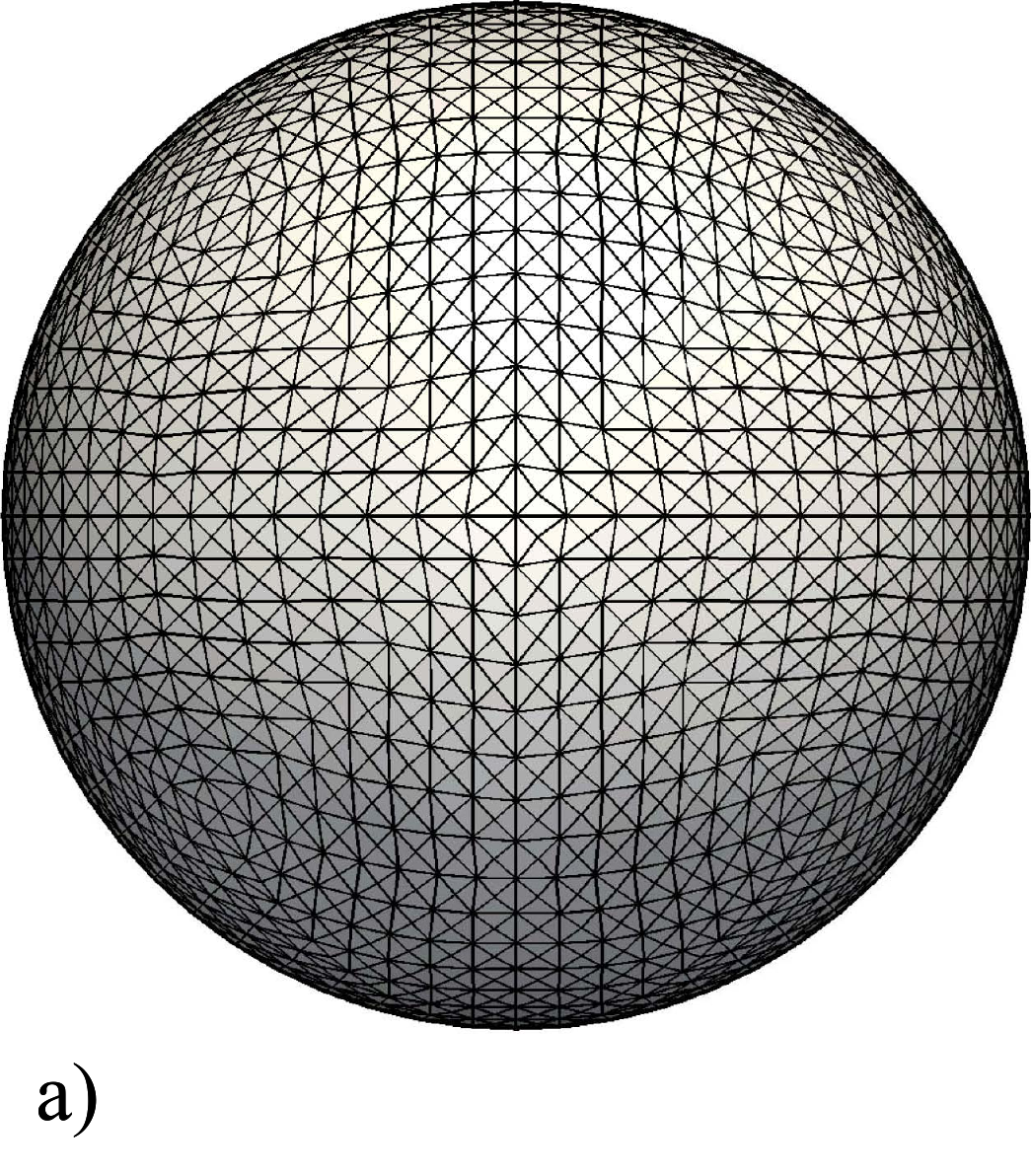}
    &
    \includegraphics[width=0.525\columnwidth]{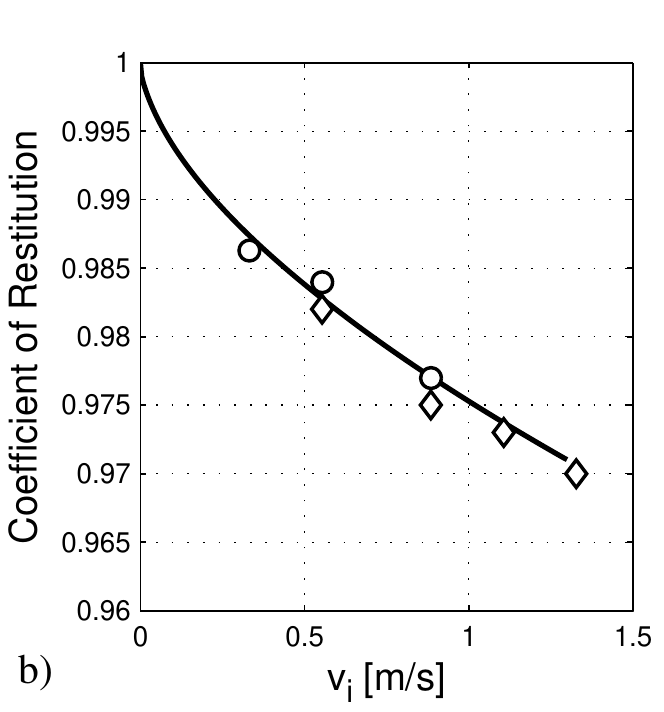}
    \end{tabular}
}
\caption{(a) Finite-element mesh of a single spherical particle. (b) Coefficient of restitution $e(v_\mathrm{i})=1-0.0247 v_\mathrm{i}^{0.61}$ obtained from best fitting (solid curve) a numerical campaign of head-on impacts of $9.525$~mm stainless steel beads (symbols). Five $v_\mathrm{i}$ are employed and $e$ is computed from $\frac{\text{KE}_f}{\text{KE}_0}=\frac{1}{2}(1+e^2)$ where $\text{KE}_0$ and $\text{KE}_f$ are the initial and final kinetic energies of the center-of-mass motion obtained from the three-dimensional finite-element simulations. Two uniform tetrahedral meshes are employed to assess convergence: (i) 53k nodes, 275k elements ($\circ$ symbols), (ii) 410k nodes, 2.2M elements ($\diamond$ symbols), per spherical particle.The mesh referred as (i) is depicted in (a).}
\label{Fig-1D-CR}
\end{figure}

\begin{figure*}[t]
\centering{
\begin{tabular}{c}
    \includegraphics[scale=0.53]{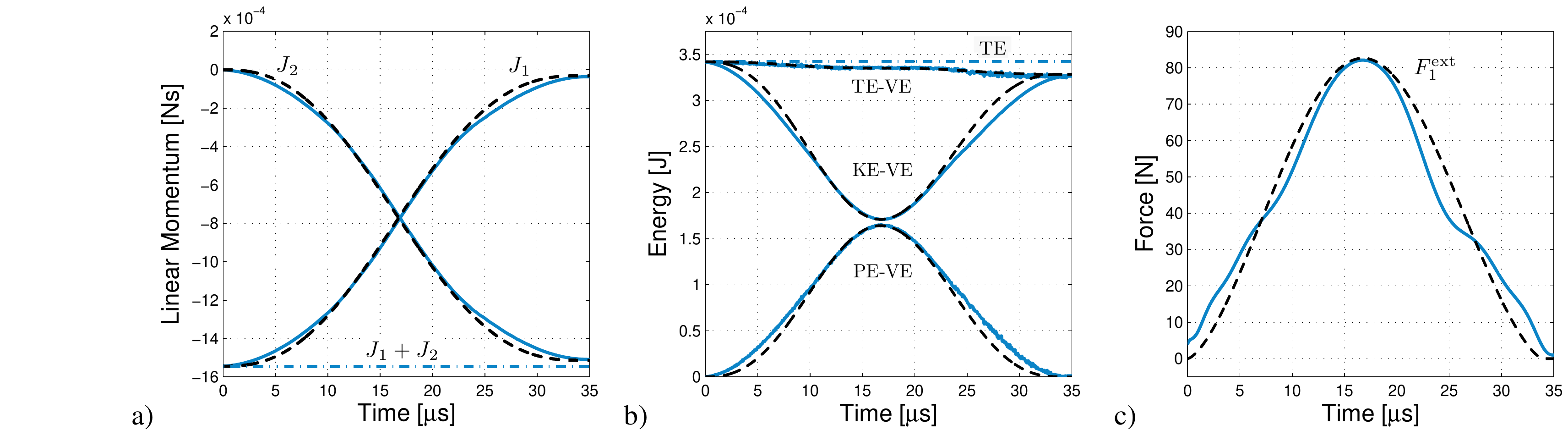}
    \\
    \includegraphics[scale=0.53]{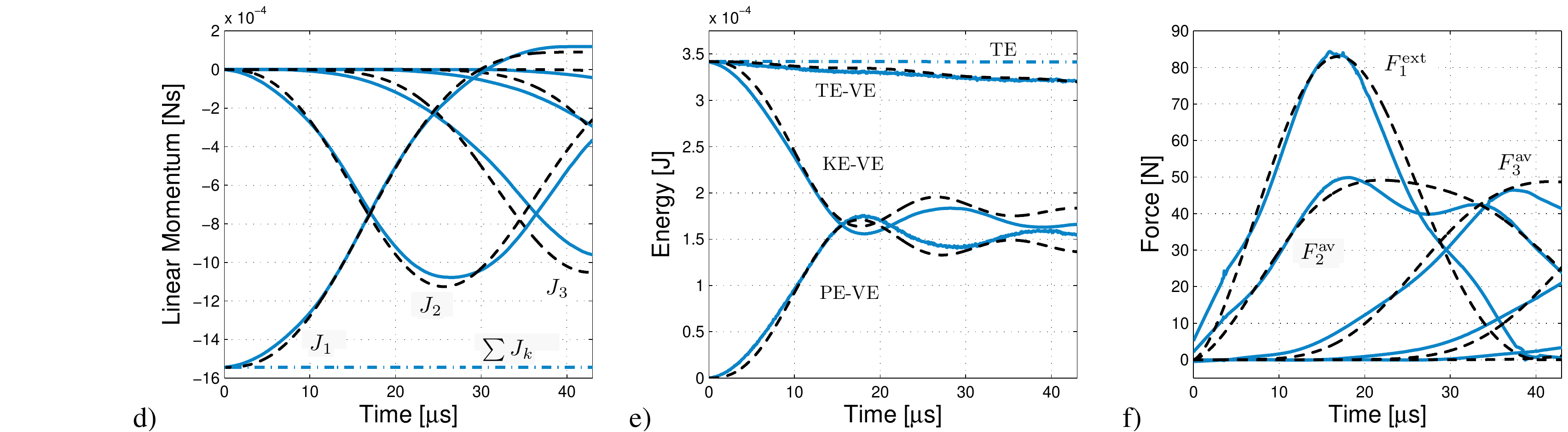}
\end{tabular}
}
\vspace{-0.1in}
\caption{(Color online) Dynamics of one bead (a,b,c) and of a one-dimensional chain (d,e,f) impacted by an identical $9.525$~mm stainless steel bead. Predictions of the 1D (3D) model are depicted by dashed (solid) curves. Energies of the 3D model are effectively compared by subtracting the vibrational components explicitly computed with the model (b,e). Exact conservation of linear momentum in both models (a,d) and total energy in the 3D model (b,e) are depicted by dash-dot curves.}
\label{Fig-Verification}
\end{figure*}

The verification of the models is twofold: (a) the assessment of the convergence and accuracy of the numerical solutions to the exact solutions of the models; and (b) the assessment of the accuracy of the one-dimensional regularized contact model as an approximation of the three-dimensional finite-element model. The former is formally addressed in \cite{Gonzalez-2010a,Gonzalez-2010b,Kane-2000}, the latter is addressed by numerical experimentation in this work. Specifically, we study the dynamics of one bead, and a one-dimensional chain, impacted by an identical $9.525$~mm stainless steel bead with $v_{\text{imp}}=0.44$~m/s. A uniform finite-element mesh with 275k 4-node tetrahedral isoparametric elements and 53k nodes per spherical particle is employed [see Fig.~\ref{Fig-1D-CR}(a)]. The coefficients of restitution $e(v_\mathrm{i})$ required in the simulations are determined from a numerical campaign of head-on impacts with the three-dimensional model. Figure~\ref{Fig-1D-CR}(b) describes the procedure followed to determine $e(v_\mathrm{i})$.

As indicated above, the regularized contact model is designed to exactly predict the coefficient of restitution of a single collision, as a consequence of solving Gothier's model (\ref{Eqn-Gonthier-1}). However, an accurate prediction of the dynamic behavior of the system is required to successfully apply the model to granular dynamics. Thus, we compare the three-dimensional and the one-dimensional models in their predictions of the evolution of: (a) the linear momentum of each particle $J_k$ obtained from the three-dimensional (respectively, one-dimensional) model as $J_k=M_k \langle\dot{\mathbf{q}}\rangle_k \cdot \hat{\mathbf{n}}$ (respectively, $J_k=M_k\dot{u}_k$); (b) the kinetic (KE), potential (PE), and total (TE) energies; (c) the external $F_k^\mathrm{ext}$ and averaged $F_k^\mathrm{av}$ forces. The vibrational energy of the three-dimensional model, i.e., $\sum\mathrm{VKE}_k$, is explicitly computed and subtracted from the system in order to compare the predictions with those of the one-dimensional model.

Figure~\ref{Fig-Verification} presents the results of the verification. The accord between the evolution in time of the total energy of the one-dimensional model and the total energy of the three-dimensional model \textit{adjusted} by the vibrational energy retained in the system is noteworthy [Figs.~\ref{Fig-Verification}(b) and \ref{Fig-Verification}(e)]. It also bears emphasis the good agreement in the averaged contact force amplitude $F_\mathrm{m}$ \cite{Daraio-2005a} and contact duration [Figs.~\ref{Fig-Verification}(c) and \ref{Fig-Verification}(f)]. Additionally, these numerical results showcase the exact conservation of linear momentum in both models [Figs.~\ref{Fig-Verification}(a) and \ref{Fig-Verification}(d)] and total energy in the three-dimensional model [Figs.~\ref{Fig-Verification}(b) and \ref{Fig-Verification}(e)].

\section{Validation}

\begin{figure}[t]
\centering{
\begin{tabular}{c}
    \includegraphics[scale=0.53]{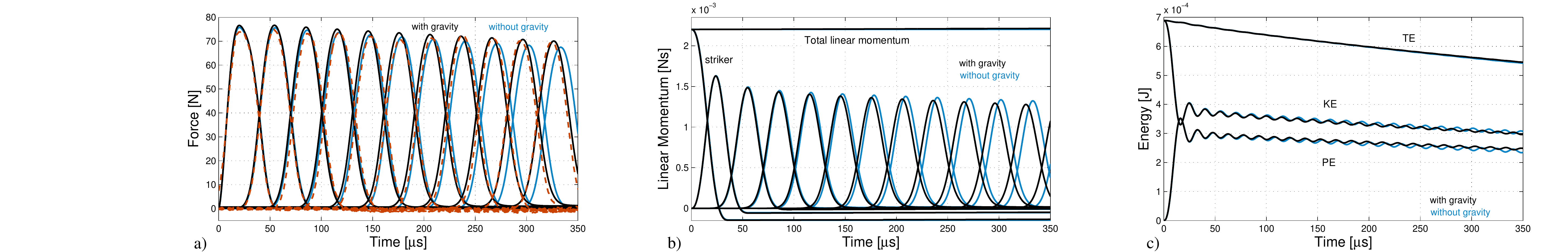}
    \\
    \includegraphics[scale=0.53]{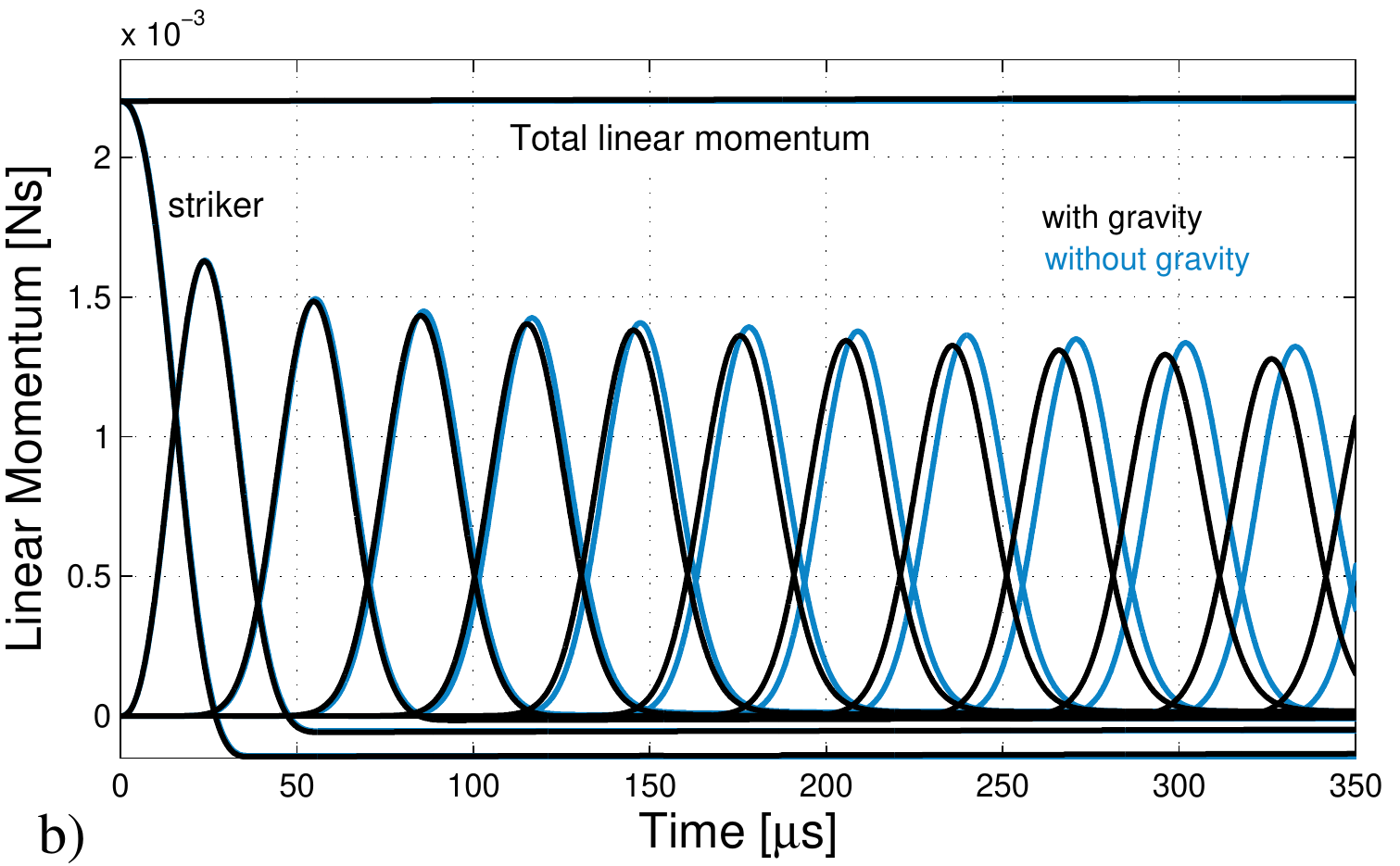}
    \\
    \includegraphics[scale=0.53]{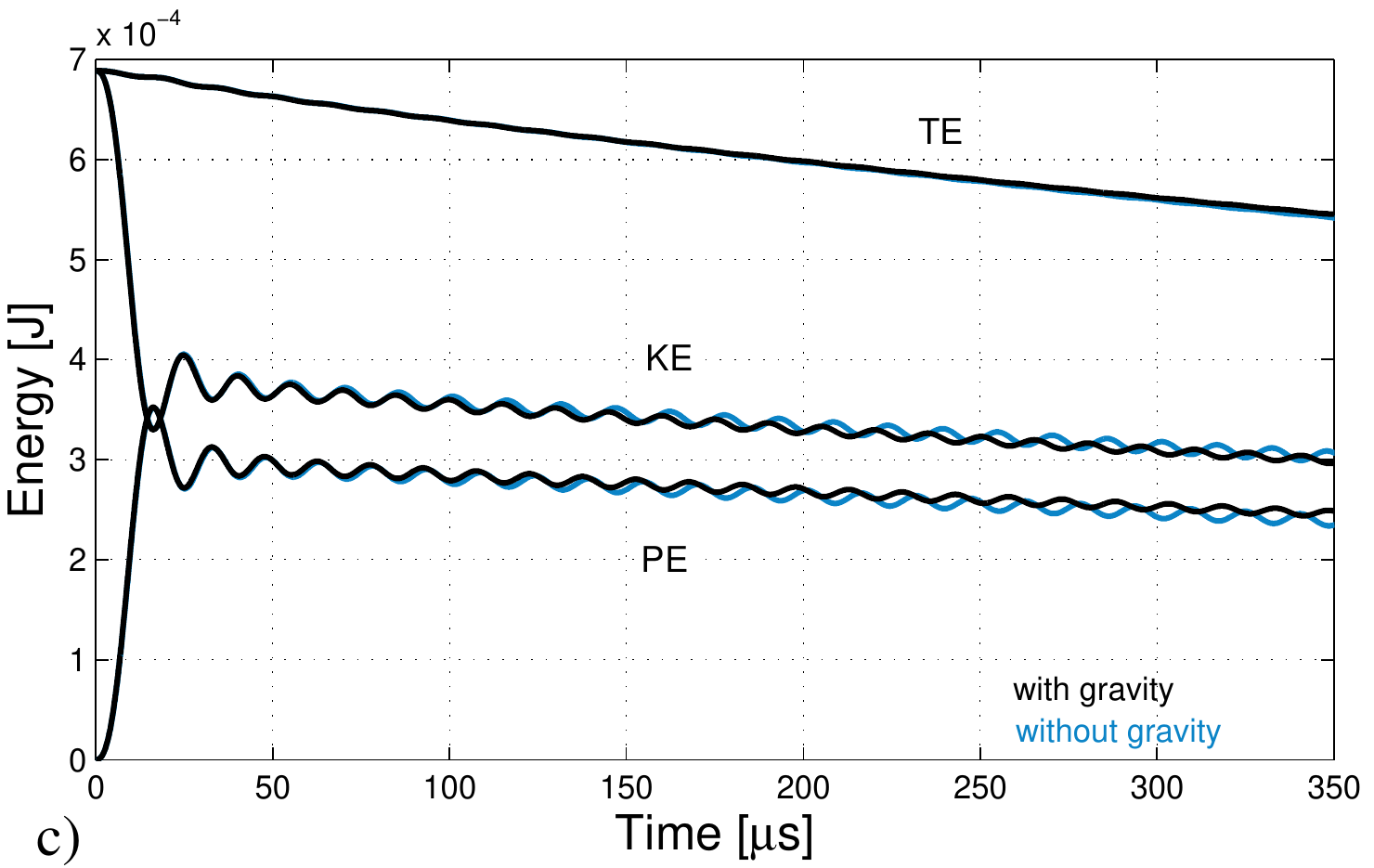}
\end{tabular}
}
\caption{(Color online) One-dimensional regularized contact model for $v_{\text{imp}}=0.63$~m/s. (a) Time history of measured forces (dashed curves) and numerical predictions with and without gravitational forces. (b) Conservation of total linear momentum and time history of individual $J_k$. (c) Time history of total (TE), kinetic (KE), and potential (PE) energies.}
\label{Fig-1D-ExpVSNum}
\end{figure}

For validation purposes, we use a set of experimental observations for a one-dimensional vertical chain of 50 stainless steel beads with diameter $9.525$~mm, impacted by an identical stainless steel bead with several $v_{\text{imp}}$. The material properties of the beads are $E=200$~GPa, $\nu=0.3$, and the density is $\rho = 7900$~kg/m$^3$. The time history of forces measured by calibrated piezosensors inserted in even-numbered beads is presented in Fig.~\ref{Fig-1D-ExpVSNum}(a)---the bead impacted by the striker is referred as the first bead. Characteristically, the measured force decays as the solitary wave propagates along the chain.

\begin{figure}[t]
\centering{
\begin{tabular}{c}
    \includegraphics[scale=0.53]{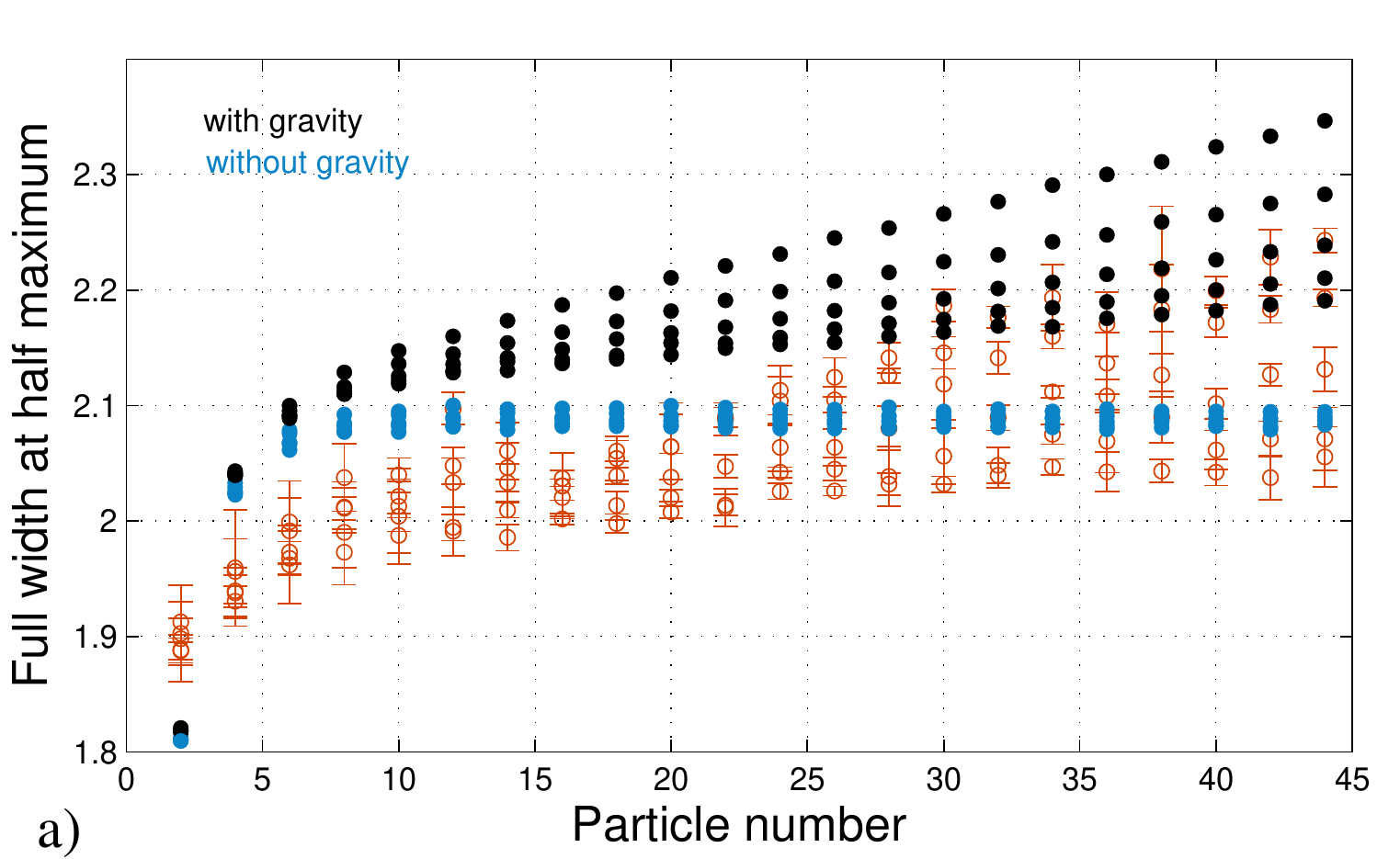}
    \\
    \includegraphics[scale=0.53]{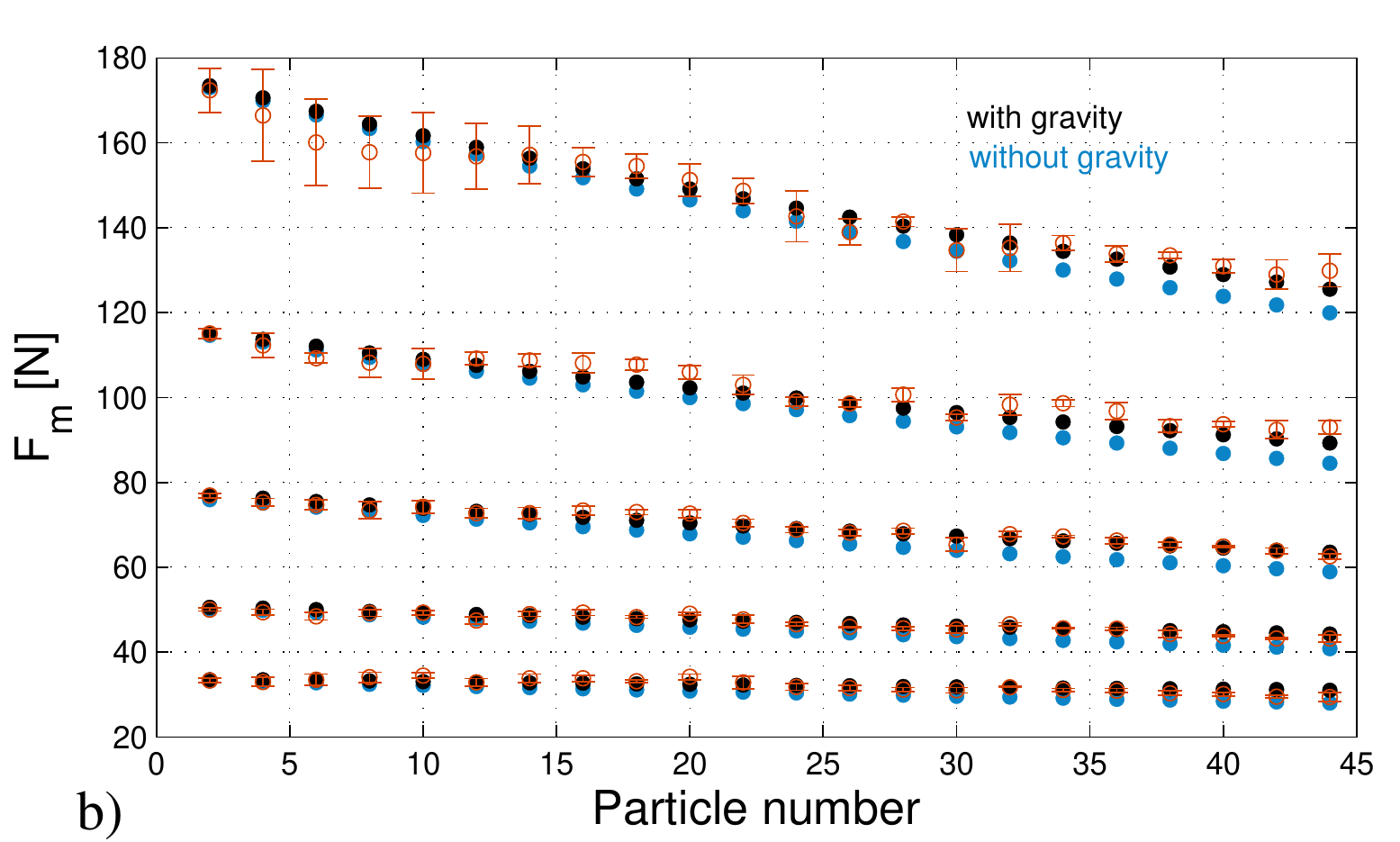}
    \\
    \includegraphics[scale=0.53]{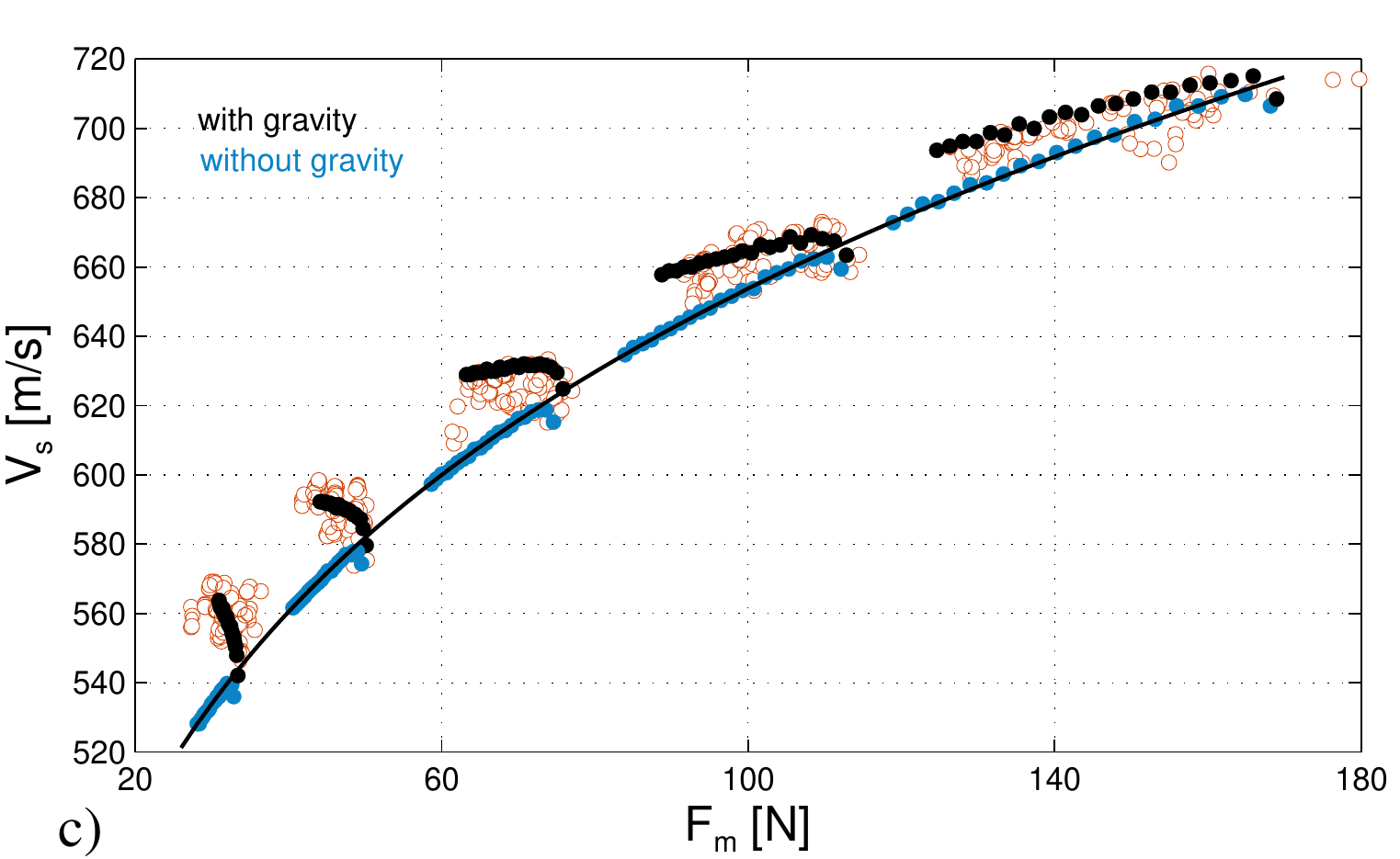}
\end{tabular}
}
\caption{(Color online) Comparison of predictions obtained with our one-dimensional numerical model, with and without gravitational forces, with measured values (empty dots) for all five $v_{\text{imp}}$. (a) Evolution of the solitary wave full width at half maximum as a function of the particles position in the chain. (b) Decay of the maximum force. (c) Speed of the solitary wave versus the maximum force and its best fit $V_s \propto F_\mathrm{m}^{0.168}$. Five different $v_{\text{imp}}$ are used, $0.31$, $0.44$, $0.63$, $0.89$ and $1.25$~m/s. Error bars represent the standard deviation from five samples of each data point.}
\label{Fig-1D-Results2}
\end{figure}

Figure~\ref{Fig-1D-ExpVSNum}(a) shows the time histories of the forces predicted by the one-dimensional regularized contact model with and without gravitational effects, i.e., by equations of motion (\ref{Eqn-MotionForcingGravity}) and (\ref{Eqn-MotionForcing}), respectively. The good agreement with the experimental observations is evident in the figure. The total linear momentum of the system without gravitational forces is a constant of motion, and this momentum-preserving property of the model is verified in Fig.~\ref{Fig-1D-ExpVSNum}(b). The decaying behavior in the total energy of the system---that resembles the vibrational energy trapped in the beads---is observed in Fig.~\ref{Fig-1D-ExpVSNum}(c). It is worth noting that both models presented in this work are intertwined in the validation since $e(v_\mathrm{i})$ is obtained from three-dimensional finite-element simulations.

It is also interesting to note that the mesoscopic approach predicts all the well-known qualitative behavior of one-dimensional granular crystal dynamics, with and without the addition of the nonuniform gravitational preload. Particularly, we observe in Fig.~\ref{Fig-1D-Results2}: (a) the formation and propagation of solitary waves with a finite width (measured as full width at half maximum) that follows the experimental trend and that is independent of $F_\mathrm{m}$, Fig.~\ref{Fig-1D-Results2}(a); (b) a decay of the force that follows $F_\mathrm{m} \propto \text{e}^{-\eta k}$, Fig.~\ref{Fig-1D-Results2}(b); (c) a solitary wave speed that follows the experimentally-observed trend and $V_s \propto F_\mathrm{m}^{0.168}$ Fig.~\ref{Fig-1D-Results2}(c). Evidently, gravitational effects are not negligible for the length of the chain and for the impact velocities considered in this work. Most importantly, vibrational-energy trapping effects appear to play a relevant role in the dynamic behavior of one-dimensional granular crystals under small to moderate impact velocities, and they explain the experimentally-observed decay behavior of $F_\mathrm{m}$.

\section{Summary and conclusions}

We have presented a mesoscopic approach to granular crystal dynamics, which comprises a three-dimensional finite-element model and a one-dimensional regularized contact model. The approach aims at investigating the role of vibrational-energy trapping effects in the overall dynamic behavior of one-dimensional chains under small to moderate impact velocities. The three-dimensional finite-element model resolves the fine mesoscale structure of dynamic collisions and explicitly accounts for the vibrational kinetic energy retained in each bead as the solitary wave propagates along the chain. The one-dimensional regularized contact model accounts for mesoscopic dynamic effects by means of a restitution coefficient, i.e., the vibrational energy that remains trapped after impact is subsumed under the concept of a coefficient of restitution. We have specifically adopted the compliant normal-force model proposed by Hunt and Crossley, and the damping-factor model proposed by Gonthier and co-workers. The only inputs of these models are the geometry and the elastic material properties of the individual particles that form the granular crystal (e.g., we have obtained the variation of the coefficient of restitution with the impact velocity from a numerical campaign of head-on collisions).

We have also presented a detailed verification and validation of the mesoscopic approach, which include: (a) the assessment of the accuracy of the one-dimensional regularized contact model as an approximation of the three-dimensional finite-element model; (b) the one-to-one comparison of the experimental and simulated time histories of averaged forces in a one-dimensional chain of 50 stainless steel beads impacted at five different velocities covering the range from $0.31$ to $1.25$~m/s. The good agreement of the later and the ability of the model to predict well-known properties of one-dimensional granular crystal dynamics (e.g., the formation of solitary waves with a finite width that is independent of the solitary wave amplitude, the decay of the force as the solitary wave propagates along the chain, and the scaling of the solitary wave speed with the wave amplitude) are noteworthy.

Finally, we remark that the mesoscopic approach predicts the experimentally-observed decay of the solitary wave amplitude from first-principles for the first time by solely addressing the role of vibrational-energy trapping effects. We thus conclude that these effects play a central role in the dynamic behavior of one-dimensional granular crystals under small to moderate impact velocities. The extension of this approach to multi-dimensional granular crystals is a worthwhile direction of future research.

\section*{Acknowledgments}

MG and MO acknowledge portions of this work funded by the DoE NNSA under Award DE-FC52-08NA28613. JY and CD acknowledge the US NSF under Award CMMI 0844540-CAREER and the ARO under Awards 54272-EG, Dr. Bruce LaMattina and MURI, Dr. David Stepp.


\begin{thebibliography}{9}
\bibitem {Sln-Alpha} A valid solution for $\alpha$ has to satisfy $\gamma^n + \alpha \gamma^n \dot{\gamma} \geq 0$.
\end{thebibliography}
\end{document}